\documentclass[prl,twocolumn,superscriptaddress,amsmath,amsfonts,
               floatfix]{revtex4}

\usepackage{epsfig,psfrag}
\usepackage{CJK}

\usepackage{graphicx}
\usepackage{dcolumn}
\usepackage{bm}
\usepackage{amssymb}
\usepackage{natbib}
\usepackage{upgreek}
\usepackage{longtable}

\begin{document}
\title{The rotating excitons in two-dimensional materials: Valley Zeeman effect and chirality}
\author{Yu Cui}
\affiliation{Tianjin Key Laboratory of Low Dimensional Materials Physics and Preparing Technology, Department of Physics, School of Science, Tianjin University, Tianjin 300354,  China}
\author{Xin-Jun Ma}
\affiliation{Research Team of Extreme Condition Physics, College of Physics and Electronic Information,
Inner Mongolia Minzu University, Tongliao 028043, China}
\author{Jia-Pei Deng}
\affiliation{Tianjin Key Laboratory of Low Dimensional Materials Physics and Preparing Technology, Department of Physics, School of Science, Tianjin University, Tianjin 300354, China}
\author{Shao-Juan Li}
\affiliation{State Key Laboratory of Luminescence and Applications, Changchun Institute of Optics,
Fine Mechanics and Physics Chinese Academy of Sciences, Changchun 130033, China}
\author{Ran-Bo Yang}
\affiliation{Tianjin Key Laboratory of Low Dimensional Materials Physics and Preparing Technology, Department of Physics, School of Science, Tianjin University, Tianjin 300354, China}
\author{Zhi-Qing Li}
\affiliation{Tianjin Key Laboratory of Low Dimensional Materials Physics and Preparing Technology, Department of Physics, School of Science, Tianjin University, Tianjin 300354, China}
\author{Zi-Wu Wang}
\email{wangziwu@tju.edu.cn}
\affiliation{Tianjin Key Laboratory of Low Dimensional Materials Physics and Preparing Technology, Department of Physics, School of Science, Tianjin University, Tianjin 300354, China}

\begin{abstract}
We propose the rotational dynamics of the intralayer and interlayer excitons with their inherent momenta of inertia in the monolayer and bilayer transition metal dichalcogenides, respectively, where the new chirality of exciton is endowed by the rotational angular momentum, namely, the formations of left- and right-handed excitons at the +K and -K valleys, respectively. We find that angular momenta exchange between excitons and its surrounding phononic bath result in the large fluctuation of the effective $\textsl{g}$-factor and the asymmetry of valley Zeeman splitting observed in most recently experiments, both of which sensitively depend on the magnetic moments provided by the phononic environment. This rotating exciton model not only proposes a new controllable knob in valleytronics, but opens the door to explore the angular momentum exchange of the chiral quasiparticles with the many-body environment.
\end{abstract}
\maketitle

$Introduction.$-----A series of extraordinary optical and electric properties in transition metal dichalcogenides (TMDs) are dominated by excitons in different spin, valley and layer configurations\cite{w1,w2,w3,w5,w6,w7}. In particular, both intralayer excitons in monolayer and interlayer excitons in bilayer structures are endowed with a valley degree of freedom (valley pseudospin) at two inequivalent but energy-degenerate $\pm$K valleys\cite{w11,w12}. This valley pseudospin, like real spin, is associated with valley magnetic moment, giving rise to exciton Zeeman splitting in the presence of the external magnetic field\cite{w13,w14,w15,w16}, which provides not only an attracted method of breaking the valley degeneracy, but a powerful lever to exploit the fundamental physical properties of the valley states, as well as for the development of new approaches to valleytronic control\cite{w17,w18}.

\begin{figure}[t!]
\includegraphics[width=3.3in,keepaspectratio]{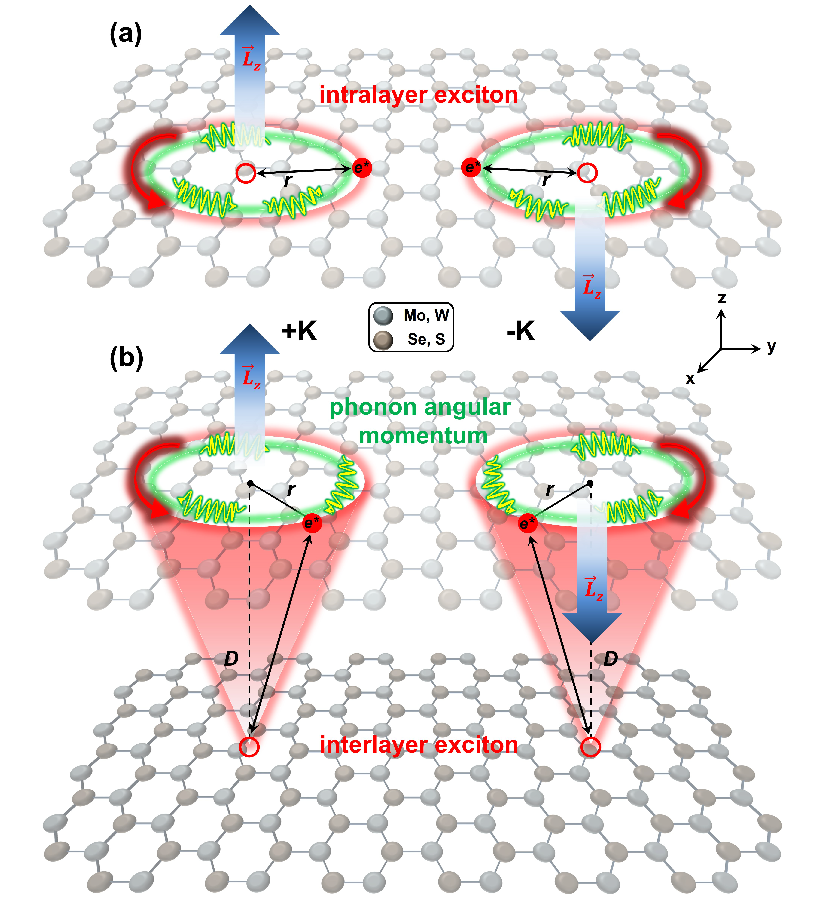}
\caption{\label{compare} (a) Schematic diagrams for the left- and right-handed rotating intralayer excitons in monolayer transition metal dichalcogenides (TMDs) at the +K and -K valleys, respectively, where an out-of-plane rotational angular momentum ${\hat {\bf{L}}_z}$ with opposite signs is formed. For a rotating exciton, the electron and hole induce the opposite magnetic moments, but they can't cancel each other out due to the difference of effective masses between them. Thus, a finite magnetic moment is produced, which could be equivalently regarded as an effective charge $e^*=e(m_h-m_e)/(m_h+m_e)$ rotating around a fixed axis with the radius $r$ (the relations between this effective charge and the valley Zeeman term are given in supplemental materials Part II). Similar to intralayer excitons, two rotating interlayer excitons in bilayer TMDs are schemed in (b). $D$ is the interlayer distance between two hosting layers.}
\end{figure}

The magnitude of this valley Zeeman splitting can be characterized by the effective $\textsl{g}$-factor ($\textsl{g}_{eff}$) of exciton, whose predicted value is -4 (or +4) including the contribution from the intra- and intercellular orbital magnetic moments of the band structures\cite{w19,w20}. However, there exists obviously disagreements between the theoretical prediction and the experimental measurement. To explain this discrepancy, several microscopic models have been proposed, e.g., the interplays among the magnetic moment of the transition metal $d$$_{x^2-y^2}$$\pm i$$d$$_{xy}$ orbitals\cite{w21,w25}, the valley magnetic moment and the lattice contribution stemming from Berry curvature should be considered\cite{we1,we2,we3}; the strain induces a hybridization of the direct and indirect excitons, giving rise to the renormalization of $\textsl{g}_{eff}$\cite{t16}; instead of the local approximation, the distinct reduction of $\textsl{g}_{eff}$ could be obtained when full Bloch states are included based on the first-principles calculation in the Bethe-Salpeter equation\cite{t17}. For the larger $\textsl{g}_{eff}$ of interlayer exciton, the mechanism that the brightening of forbidden optical transitions induced by the moir\'e potential has also been suggested\cite{w29}. Nevertheless, the accuracy of the developed models is rather poor comparing with the experimental data. The underlying physics for this factor remains a subject of hot debate.

In this letter, we investigate the rotational motion of the intralayer and interlayer excitons with their inherent momenta of inertia in monolayer and bilayer TMDs, respectively. This rotational degree of freedom gives rise to an additional orbital angular momentum in the out-of-plane due to the difference of effective masses between electron and hole, adding a finite magnetic moment that enhances the exciton Zeeman splitting upon interaction with an external magnetic field. This picture enables us, on the one hand, to define the new chirality of the exciton by the left- and right-handed
rotations at the +K and -K valleys, respectively, as shown in Fig. 1; on the other hand, to explore angular momenta transfer between excitons and phononic bath. We find that these values of $\textsl{g}_{eff}$ predicted by this model are in excellent agreement with experimental measurements for both intralayer and interlayer excitons. More importantly, the clockwise and anti-clockwise motions of the phononic bath could be used to reveal an unexplained puzzle that the asymmetry of valley splitting.

$The$ $Hamiltonian$ $of$ $the$ $rotating$ $exciton.$-----We propose an exciton as an electric dipole with the rotational motion before its recombination schemed in Fig. 1. For an exciton in two-dimensional material, its rotational degree of freedom confers upon an out-of-plane orbital angular momentum with opposite signs depending on the rotational directions\cite{t17,t20,w20}. Meanwhile, the rotating exciton is immersed into the phononic bath, leading to the exchange of angular momenta between them. In the presence of an external magnetic field, the total Hamiltonian is given by\cite{t6,t7,t8,supp}
\begin{eqnarray}
&&\mathcal{\hat H}
=\hbar\xi_0{{\hat {\bf{L}}_z^2}}+{\mu _B}{\textsl{g}_{\kappa}}\hat {\bf{B}} \cdot {\hat {\bf{L}}_z}-{\mu _B}{\textsl{g}_{0}}\hat {\bf{B}} \cdot {\hat {\bf{L}}^0_z}\nonumber\\
&&+\sum\limits_{q\lambda } {\overline {\hbar {\omega  }}_{\nu}} \hat b_{q\lambda }^\dag {\hat b_{q\lambda }}+\sum\limits_{q\lambda } {{\mathrm{V}}}_{\lambda} (q,r)\left[ {{e^{i\lambda \hat \varphi }}{{\hat b}_{q\lambda }} + {e^{ - i\lambda \hat \varphi }}\hat b_{q\lambda }^\dag } \right],\nonumber\\
\end{eqnarray}
where the first term is the kinetic energy of a rotating exciton with the rotational constant $\xi_0=\hbar/({{2I}})$, $I=\eta r^2$ is the moment of inertia depending on the reduced mass $\eta^{-1}=m_e^{-1}+m_h^{-1}$ and the relative distance $r$ between the electron and the hole; $m_e$ ($m_h$) is the electron (hole) effective mass; ${\hat L_z} =  - i\partial /\partial \varphi $ is the angular momentum operator with the eigenvalues $l_z=0,\pm1,\pm2\ldots$ and  eigenenergies $E_{l_z}=\hbar\xi_0l_z^2$\cite{t9}, where the sign $\pm$ represents two opposite directions of this angular momentum in the out-of-plane of two-dimensional materials, corresponding to the left- and right-handed excitons at the +K and -K, respectively, and representing a new chirality of exciton, which is also corroborate with the formation of valley excitons at the +K and -K valleys upon absorption of left- and right-handed circularly polarized light\cite{bbb1,bbb2}. Just this angular momentum endows a magnetic moment that results in the interaction between the rotating exciton and the external magnetic field, thus giving rise to the additional valley Zeeman effect, which can be described by the second term ${\mu _B}{\textsl{g}_{\kappa}}\hat {\bf{B}} \cdot {\hat {\bf{L}}_z}$ (the detailed derivation for it in supplemental materials Part I) with $\mu_B=e\hbar/(2m_0c)$ is the Bohr magneton and $\textsl{g}_{\kappa} = {m_0}({m_h} - {m_e})/(m_em_h)$ is the gyromagnetic ratio for the rotating exciton ($m_0$ is the free electron mass) and $\hat {\bf{B}}$ is the vector of magnetic field. In other word, the magnetic moment of the rotating exciton is equivalent to a single particle rotating around the fixed axis with the radius $r$ and the effective charge $e^*=e(m_h-m_e)/(m_h+m_e)$; see the supplemental materials Part II for details. We will see, in fact, this magnetic moment is very tiny for intralayer exciton because of the small difference of effective masses between electron and hole in TMDs. For interlayer exciton, however, it plays the key role to modulate the valley Zeeman splitting because of the tunability of the ratio of effective masses between electron and hole, hosting two dividual layers, respectively. The third term represents the intrinsic exciton valley splitting with the Lande factor $\textsl{g}_0=1$, which mainly attributes to the contribution of  $d$-orbital of transitional-metal atom with the magnetic quantum $\mathfrak{L}_z^0=+2$ at the +K valley and $\mathfrak{L}_z^0=-2$ at the -K valley, yield the ideal values $\textsl{g}_{eff}^0=-4$, which has been predicted by several theoretical models and proved by some experiments\cite{w19,w20}.
The fourth term describes the kinetic energy of the phononic bath with
$\overline {\hbar {\omega  }}_{\nu}= \sqrt {\hbar \omega _{\nu} ^j\hbar \omega _{\nu} ^{j'}}$ being the phonon energy depending on the hosting layer index $j$ $(j')$ for hole (electron). For the intralayer exciton, the same layer index $j=j'$ is fulfilled. For the interlayer exciton in bilayer TMDs, the electron and hole reside in different layers, so $j=j'$ and $j\neq j'$ correspond to the homobilayer and heterobilayer structures, respectively. Here ${q}=|{\bf q}|$ is the scalar representation of the phonon wave vector, satisfying the relation $\sum\nolimits_q { \equiv \int {dq} }$. $\lambda$ represents the phonon angular momentum.
The corresponding creation and annihilation operators, $\hat b_{\bf{q}}^\dag$ and ${\hat b}_{\bf{q}}$, are expressed in polar coordinates, $\hat b_{q\lambda}^\dag$ and ${\hat b_{q\lambda }}$\cite{tt1,supp}. The last term describes the interaction between the rotating exciton and phonons. The angular momentum-dependent coupling strength $\mathrm{V}_{\lambda} (q,r)$ is given by
\begin{eqnarray}
{{V}}_\lambda (q,r)=\sqrt {\frac{q}{{2\pi }}} \left[\mathcal{M}_h^j {{J_\lambda }\left( { - {\beta _1}qr} \right) - \mathcal{M}_e^{j'}{J_\lambda }\left( {  {\beta _2}qr} \right)} \right],
\end{eqnarray}
depending on the microscopic details of two-body interaction ${\mathcal{M}^{j(j')}_{h(e)}}=[ {{{e^2}\alpha Z_0\hbar {\omega^{j(j')} _{\nu}}}/({{2\mathbb{A}{\varepsilon _0}}})}]^{1/2}$ between hole (electron) and the $\nu$th branch of phonon mode\cite{t2,t3}, where $Z_0$ is the monolayer thickness, $\mathbb{A}$ is the quantization area in the monolayer materials, and $\varepsilon _0$ is the permittivity of vacuum. $\alpha$ denotes the strength of the rotating exciton coupled with its surrounding phononic bath and is regarded as a changeable parameter in the large scale, because of the following reasons: (i) different branches of phonon modes will give the coupling contribution together; (ii) the strength could be tuned significantly by the dielectric environment of the monolayer and bilayer structures, e.g., the various choices of the substrate and encapsulation materials, and the structural parameters, e.g., the internal distances between the hosting monolayer and the substrates\cite{t2,t3,cp2}; (iii) the moir\'e potential induced by stacking angle not only gives the new phonon modes, but enhances the exciton-phonon coupling remarkably due to the strong quantum confinement effect\cite{bbb3,bbb4}. ${J_\lambda }(-{\beta_1} qr)$ and ${J_\lambda }({\beta_2} qr)$ are the Bessel functions of the first kind, where $\beta_1=m_h/m$ and $\beta_2=m_e/m$ for hole and electron, respectively ($m$ is the total mass.).

To show the significant influence of phonon angular momenta on the rotating particle, a variational ansatz for an exciton rotating in the phononic bath is introduced based on the angulon model developed by
Schmidt and Lemeshko\cite{t6,tt1}:
\begin{eqnarray}
\left| \Psi  \right\rangle  = \sqrt C \left| j_z \right\rangle \left| 0 \right\rangle  + \sum\limits_{q\lambda } {{\beta _{q\lambda }}} \left| {j_z-\lambda} \right\rangle \hat b_{q\lambda }^\dag \left| 0 \right\rangle,
\end{eqnarray}
where ${\hat J_z} = {\hat L_z} + {\hat \Lambda _z}$ is the total angular momentum operator of the system with the eigenvalues of $j_z=0,\pm1,\pm2,\ldots$, and ${\hat \Lambda _z}$ is the collective angular momentum operator of the phononic bath. $\sqrt C$ and ${\beta _{q\lambda }}$ are the variational parameters with the normalization condition $\left| C \right| + \sum\nolimits_{q\lambda } {{{\left| {{\beta _{q\lambda }}} \right|}^2}}  = 1$, and $\left| 0 \right\rangle$ represents the vacuum of phonons. Here, the angular momentum conservation satisfies $l_z+\lambda=j_z$. By minimizing of the function $\left\langle \Psi  \right|\mathcal{\hat H} - E\left| \Psi  \right\rangle $ with respect to the parameters
$\sqrt C^*$ and ${\beta^* _{q\lambda }}$, we get the valley-dependence eigenenergies of the system (the detailed mathematical processes are given in the supplemental materials Part III)
\begin{eqnarray}
E_{\pm \mathrm{K}} &=& {\hbar\xi _0}{l_z^2}+{\mu _B}\textsl{g}^*_\kappa B l_z^{\pm \mathrm{K}}-{\mu _B}\textsl{g}_0 B \mathfrak{L}_z^{0,{\pm \mathrm{K}}}\nonumber\\
&&+{{\overline {\hbar \omega } }_{\nu}}- \sum\nolimits^{(1)}_{l_z} {(E)},
\end{eqnarray}
with
\begin{eqnarray}
{\rm{g}}_\kappa ^* = \left[1 + \sum\limits_{q\lambda } {\frac{{{\lambda{\left| {{V_\lambda }(q,r)} \right|}^2}}}{{{l_z}{{\left( {\hbar {\xi _0}j_z^2 - {{\overline {\hbar \omega } }_\nu } - \hbar {\xi _0}l_z^2} \right)}^2}}} }\right]{\rm{g}}_\kappa,\nonumber
\end{eqnarray}
\begin{eqnarray}
\sum\nolimits^{(1)}_{l_z} {(E)}=\sum\limits_{q\lambda } {\frac{{{{\left| {{V_\lambda }(q,r)} \right|}^2}}}{{{\hbar\xi _0}{j_z^2} - {{\overline {\hbar \omega } }_{\nu}} - {\hbar\xi _0}{l_z^2}}}}.\nonumber
\end{eqnarray}
Obviously, the valley Zeeman splitting is determined by the second and third terms in Eq. (4), where the energy level diagram showing the contribution to the splitting is schemed in Fig. S1 in supplemental materials\cite{supp}. The magnitude of the energy splitting with valley dependence could be rewritten as $E^S_{\pm \mathrm{K}}={\mu _B}\textsl{g}^*_\kappa B l_z^{\pm \mathrm{K}}-{\mu _B}\textsl{g}_0 B \mathfrak{L}_z^{0,{\pm \mathrm{K}}}$. In the following sections, the lowest rotational states $l_z^{+ \mathrm{K}}=+1$ ($l_z^{- \mathrm{K}}=-1$) for the +K (-K) valley is adopted.  The impact of phonon magnetic moment on the valley splitting is reflected by the renormalization of ${\rm{g}}_\kappa ^*$ in the second term. Consequently, the effective ${\textsl{g}}-$factor of exciton could be obtained by ${\textsl{g}}_{eff}=(E^S_{+ \mathrm{K}}-E^S_{- \mathrm{K}})/(\mu_BB)$. The last term is the self-energy of the rotating exciton due to the angulon effect, which, in fact, only gives the tiny contribution to the valley splitting, but plays the crucial role in inducing the fine structure of the spectroscopy for the rotating particles in the many-body environment. Here, the first-order approximation of this term is adopted (see the supplemental materials Part III).

We mainly discuss the renormalization of  ${\textsl{g}}-$factors for the rotating intralayer and interlayer excitons in four typical monolayer TMDs and their bilayers structures, where
phonon angular momenta $\lambda=-1$ and $\lambda=+1$ are selected in the calculation processes\cite{t11}, representing the clockwise and anti-clockwise rotations of the phononic bath respecting to the motion of exciton, respectively.
The adopted values of other parameters for these materials are shown in Table S1 and Table S2 in the supplemental materials\cite{supp}. In addition, we only consider the A-type excitons in this work and assume that B-type excitons have the similar behavior.

$Exciton$ $\textsl{g}_{eff}$ $in$ $monolayer$ TMDs.-----
 Fig. 2(a) shows the Zeeman splitting of intralayer excitons with the magnetic field for four well-known monolayer TMDs, where the precondition is assumed that the orientation of the phonon angular moment is always keeping opposite respecting to the right- and left-handed rotating excitons at the +K and -K valleys, respectively. Comparing to the normal valley Zeeman splitting (purple solid lines for $\textsl{g}_{eff}=-4$)\cite{w19,w20}, the slopes of energy shift decrease obviously under the influence of the transfer of phonon angular momentum, which also directly show the existence of phonon magnetic moment, and even the corresponding magnitude could be evaluated quantitatively.
\begin{figure*}
\centering
\includegraphics[width=7in,keepaspectratio]{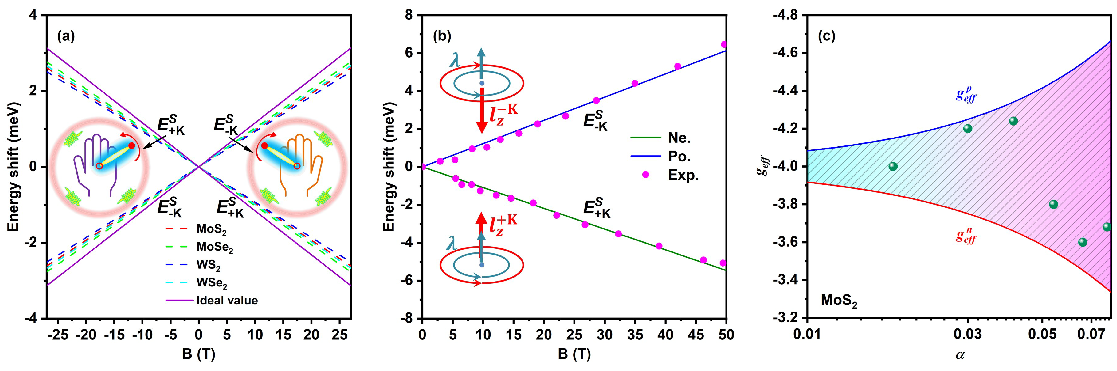}
\caption{\label{compare} (a) Energy shift of intralayer exciton at +K and -K valleys without (solid lines) and with (dash lines) the transfer of phonon angular momentum in the presence of the magnetic field at $\alpha=0.08$. (b) Energy shifts of exciton with the contribution of the phonon angular momentum $\lambda=+1$ at the +K and -K valleys for the monolayer WS$_2$ at $\alpha=0.024$, in which experimental data (pink dots) are reproduced from Ref. 20. (c) The renormalization of $\textsl{g}-$factors with the positive (the upper-limit $\textsl{g}^p_{eff}$) and negative (the lower-limit $\textsl{g}^n_{eff}$) contribution of phonon magnetic moments as a function of the coupling constant $\alpha$ in monolayer MoS$_2$, where these solid dots represents the values of $\textsl{g}_{eff}$ obtained in experiments.
}
\end{figure*}
In addition, the slopes of energy shift for the W-based monolayers are flatter than Mo-based ones. This behavior could be attributed to two reasons: the stronger strength of the exciton-phonon coupling and the bigger value of the gyromagnetic ratio ${\rm{g}}_\kappa$ stemming from the larger difference of effective masses between electron and hole in W-based monolayer (see the Table S1), both of which lead to the more correction to exciton magnetic moment.

Another abnormal phenomenon that the asymmetrical distribution of Zeeman shifts at two valleys in the monolayer TMDs has been widely observed by earlier experiments\cite{t17,we2}. To explain it from the microscopic aspect, several advanced theoretical models based on the first-principles calculation have been given, e.g., Caruso $et$ $al.$ proposed the out-of-plane component of the orbital angular momentum of an exciton is expressed in terms of the Bethe-Salpeter equation\cite{t17}, Deilmann $et$ $al.$ developed a new approach merging the contribution of full Bloch states to the excitonic magnetic moments\cite{w24}. However, the effective simulations for the experimental measurements are still lacking.  Fig. 2(b) shows the valley Zeeman shift of the rotating exciton as a function of the magnetic field in monolayer WS$_2$ (solid lines), along with magneto-reflectance spectroscopy data from Ref. 20 (pink dots), where phonon angular momentum $\lambda=+1$ at both the +K and -K valleys is assumed. Namely, the direction of phonon angular momentum is the same to the exciton angular momentum at the +K valley, but reversely at the -K valley, as illustrated in the insets. One can see that the experimental data are very excellently fitted with the negative and positive magnetic moments provided by phonon angular momentum at the +K and -K valley excitons, respectively. The quantum state $\lambda=-1$ also has the similar effect. This implies that (i) the rotation of valley exciton is not just hindered by the phononic bath, and it also can get the assistance from the surrounding environment; (ii) the magnitude of Zeeman shift could be varied in the large scale due to the phononic bath endowing the positive and negative contribution to the exciton magnetic moment in each valley, which may provide a potential explanation for the large fluctuation of $\textsl{g}_{eff}$ in experiments. To show the latter effect clearly, we present the renormalization of $\textsl{g}_{eff}$ as a function of the coupling constant for the monolayer MoS$_2$ in Fig. 2(c). According to the positive and negative contribution to the exciton magnetic moment, the upper- ($\textsl{g}^p_{eff}$) and lower-limits ($\textsl{g}^n_{eff}$) of the renormalization of $\textsl{g}_{eff}$ are presented, in which some values of $\textsl{g}_{eff}$ measured by recent experiments are also illustrated\cite{we2,w23,w24,e1,e2,e3,e4,t19}. It is obviously show that the large fluctuation of $\textsl{g}_{eff}$ could be covered successfully with increasing the coupling strength. The same results are obtained for other monolayer TMDs in Fig. S3.

From 2015, Schmidt and Lemeshko introduced a new quansiparticle---angulon, describing the quantum impurity rotating in phononic bath\cite{t6}, such as molecules merging into superfluid helium droplets. They pointed out the angulon induces the rotational fine structures due to the angular momenta transfer between the rotor and the phonons. These transfer processes reflected by the rotational Lamb shift in spectra have been proved in experiments\cite{t13,t14,t15}. However, the distinguishment between the clockwise and anti-clockwise motions for phononic bath is still a challenge task. The model of the rotating exciton in these two-dimensional TMDs systems presents an ideal platform to overcome this problem by analyzing the renormalization of the $\textsl{g}-$factor of the valley exciton.

\begin{figure*}
\centering
\includegraphics[width=7in,keepaspectratio]{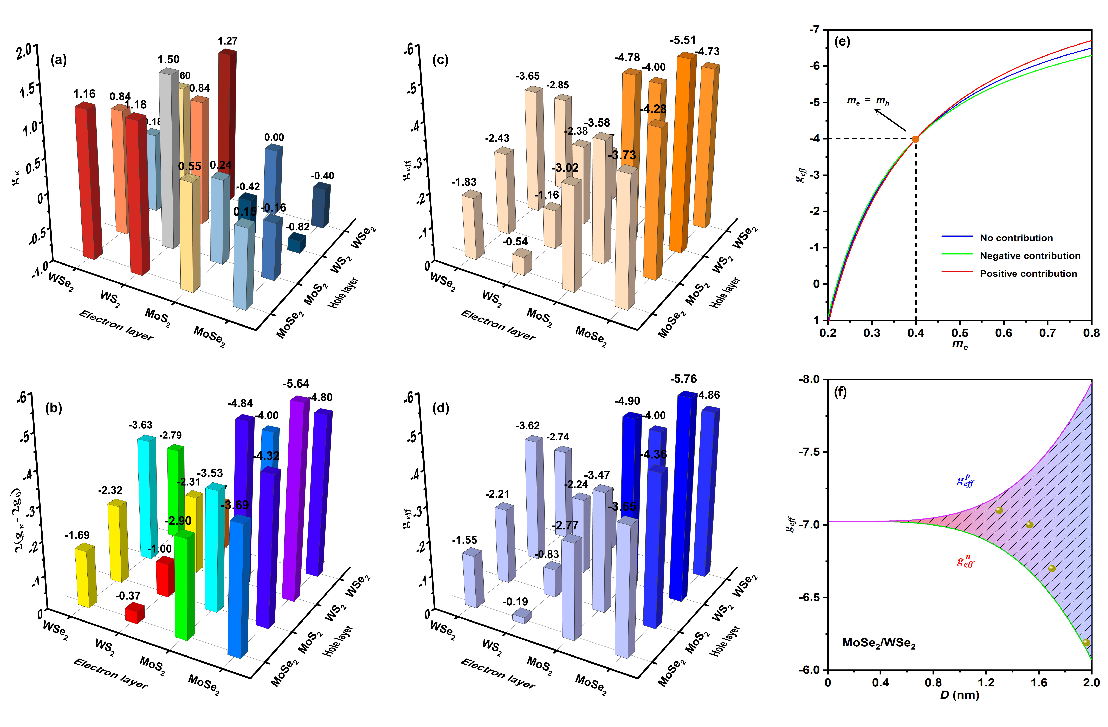}
\caption{\label{compare} (a) $\textsl{g}_\kappa$ for interlayer excitons in sixteen TMDs bilayers based on the difference of effective masses between electron and hole. (b) The correction of  $\textsl{g}^0_{eff}=-4\textsl{g}_0$ by the 2$\textsl{g}_\kappa$ for interlayer excitons in sixteen TMDs bilayers. The renormalization of $\textsl{g}_{eff}$ for the rotating interlayer excitons with the positive (c) and negative (d) phonon magnetic moments in these bilayer structures at $\alpha=0.13$, $D=1.5$ nm. (e) The impact of the effective mass ratios between electron and hole on the  $\textsl{g}_{eff}$ with and without the contribution of phonon magnetic moments. (f) $\textsl{g}_{eff}$ depends on the internal distance with the positive (the upper-limit $\textsl{g}^p_{eff}$) and negative (the lower-limit $\textsl{g}^n_{eff}$) phonon magnetic moments in MoSe$_2$/WSe$_2$ heterostructure at $\alpha=0.35$.}
\end{figure*}

$Exciton$ $\textsl{g}_{eff}$ $in$ $bilayer$ TMDs.-----Comparing with the intralayer excitons, the interlayer excitons have more tunabilities stemming from the separation of the hosting layers for electron and hole as well as the modulation of internal distance between two hosting layers. For instance, four typical monolayer TMDs could be stacked into sixteen bilayer structures, including four homostructures and twelve heterostructures. Based on the difference of effective masses between electron and hole (see Table S1 in the supplemental materials), these values of the gyromagnetic ratios ($\textsl{g}_\kappa$) for sixteen rotating excitons and their correction to the ideal value $\textsl{g}^0_{eff}=4$ are listed in Figs. 3(a) and (b), respectively. We can see that $\textsl{g}_\kappa$ varies remarkably with increasing the ratios of effective masses between electrons and holes, and thus results in the larger variation of $\textsl{g}_{eff}$ in Fig. 3(b), which presents a very practical way to control the valley Zeeman splitting of interlayer exciton by different ratios of effective masses of electron-hole pairs. Furthermore, the renormalization effect of the angular momentum transfer of the phononic bath on $\textsl{g}_{eff}$ are shown in Figs. 3(c) and (d) with the positive and negative magnetic moment contribution at $\alpha=0.13$, respectively. One can see that phonon angular momenta give the smaller correction to $\textsl{g}_{eff}$ comparing with the ratio of effective mass. Besides these monolayer TMDs materials, various of heterostructures could be composed of by the huge members of two-dimensional material family for interlayer excitons, ensuring the almost continuously changing ratios of effective mass for electron-hole pairs. This allows us to obtain much larger variation for the $\textsl{g}_{eff}$ as plotted in Fig. 3(e). Moreover, the positive and negative correction of phonon magnetic moments will be enhanced clearly with the increasing this ratio. These results suggest that the difference between electron and hole plays a predominate role to adjust the $\textsl{g}-$factor of interlayer exciton in two-dimensional van der Waals heterostructures.

Another controllable parameter for the interlayer exciton is the internal distance $D$ between two hosting layers that related to the rotational constant $\xi_0$ (see the supplemental materials Part II). The interlayer distance dependence of $\textsl{g}_{eff}$ is given in Fig. 3(f) for MoSe$_2$/WSe$_2$ bilayer with the positive ($\textsl{g}^p_{eff}$) and negative ($\textsl{g}^n_{eff}$) contribution of phonon magnetic moments. We find the renormalization of $\textsl{g}_{eff}$ varies in the range of $-4.5\sim-9.5$ with the increasing of $D$, covering most experimental data (solid dots) illustrated in Fig. 3(f)\cite{w28,w29,t19,t32,t33}. This behavior can be attributed to the enhanced transfer of phonon angular momentum with $D$, resulting in the large fluctuation of $\textsl{g}_{\kappa}^*$. Hence, the interlayer distance is also a key parameter for the modulation of the rotational interlayer exciton. But this role has been neglected in most experiments, which may explain why different values of $\textsl{g}_{eff}$ were obtained for the same bilayer structures\cite{w28,w29,t19,t32,t33}.

Recently, the chiral phonon has been predicted by the theoretical studies and proved by some experiments in two-dimensional TMDs structures\cite{ch1,ch2,ch3,ch4,ch5,ch6}, which plays an important role for the interconversion between the dark and bright excitons due to the required momentum conservation. Here, the chirality of the rotational angular momentum constitutes an inherent degree of freedom of valley excitons that induces the coupling to other degree of freedom, e.g., spin and orbital angular momentum, and the external perturbations\cite{jg1}. In turn, the chirality of other degrees of freedom could be reflected by the exchange of angular momenta between them. On the other hand, the larger fluctuation of $\textsl{g}_{eff}$ for trion has been observed in recent experiments\cite{jg2,jg3,jg4}. Using the rotating exciton model, we present a potential explanation  that an extra electron (or hole) in a rotating trion will induce the bigger magnetic moment, enhancing the valley Zeeman effect. Moreover, the transfer of phonon angular momentum results in the fine-structures of trions.

$Conclusion.$-----In summary, we propose the rotational motion of valley excitons in monolayer TMDs and their bilayer structures, where the new chirality of exciton is defined by the rotational angular momentum, which could be manifested by the valley Zeeman splitting. Furthermore, the rotating excitons induce the transfer of phonon angular momentum, resulting in the renormalization effect of exciton $\textsl{g}-$factors, and indicating the chirality of the phonon and phonon magnetic moment. These results also provide important enlightenments for revealing the underlying physics of the optical properties of valley excitons in two-dimensional materials.

This work was supported by National Natural Science Foundation of China (Grant Nos. 11674241, 62022081, 61974099 and 12174283).

\end{document}